\documentclass{article}
\usepackage{spconf,amsmath,amsfonts}
\usepackage{graphicx,booktabs,multirow,array}
\usepackage{cite}
\usepackage[colorlinks=true,linkcolor=black,citecolor=black,urlcolor=blue]{hyperref}
\begin{document}
\ninept

\title{Robust Rank Aggregation for Multimodal\\Speech-Based Alzheimer's Disease Detection}

\name{Zemin Jin$^{1}$ and Tomoko Matsui$^{2}$\sthanks{Corresponding author.}}
\address{$^{1}$The Chinese University of Hong Kong, Shenzhen, China\\
$^{2}$Center for Language, Intelligence and Machines (LIMA),\\
Shenzhen Loop Area Institute (SLAI), Shenzhen, China\\
Emails: zeminjin@link.cuhk.edu.cn, tomokomatsui@slai.edu.cn}

\maketitle

\begin{abstract}
Speech-based Alzheimer's disease (AD) detection has recently benefited from multimodal foundation-model representations that integrate complementary acoustic and linguistic information. However, conventional probability averaging over these complementary classifiers is unreliable, because their posterior probabilities exhibit mismatched scales: identical values may reflect different confidence levels across models. We propose a robust rank aggregation framework that aggregates normalized prediction ranks instead of posterior probabilities. Each subject is scored by its percentile within a \emph{fixed training-cohort} distribution of out-of-fold predictions; since rank ordering is invariant to monotonic transformations, this avoids probability-scale mismatch while preserving classifier confidence ordering. A confidence-gated Random Forest further corrects residual errors using clinically interpretable linguistic features, overriding the rank prediction only when the two disagree and the RF is highly confident, without additional deep model training or explicit posterior-probability calibration. On ADReSS2020 and ADReSSo2021, the method achieves accuracies of 95.83\% and 90.14\%, respectively, comparing favorably with previously reported results.

\end{abstract}

\begin{keywords}
Alzheimer's disease, speech analysis, multimodal ensemble, rank aggregation, random forest
\end{keywords}
\section{Introduction}
\label{sec:intro}

Alzheimer's disease (AD) is the leading cause of dementia, accounting for an estimated 60--70\% of dementia cases worldwide~\cite{who2026}. As populations age, early detection becomes an increasingly important public health goal. Conventional diagnostic procedures are often time-consuming, expensive, and limited to clinical settings, which makes large-scale screening difficult. Automatic analysis of spontaneous speech offers a practical alternative, and the ADReSS2020 and ADReSSo2021 challenges provide standardized datasets for developing and evaluating such systems~\cite{luz2020,luz2021}.

To build an effective system for these benchmarks, a natural strategy is to combine multiple models rather than rely on a single classifier, since individual models are unstable on small clinical datasets and aggregating multiple runs or models has been shown to reduce variance and improve accuracy~\cite{yuan2020,qiao2021}. Beyond this practical concern, Krogh and Vedelsby~\cite{krogh1994} further showed that combining diverse models reduces the generalization error below the average error of the individual members, with the gain proportional to the disagreement among members. In this work, we create diversity across modalities: HuBERT~\cite{hsu2021} learns acoustic representations from the waveform, while RoBERTa~\cite{liu2019} learns lexical and semantic representations from the transcript. We also create within-representation diversity by training logistic regression classifiers with different $L_2$ regularization strengths, which yields different decision boundaries and confidence scores. Multimodal systems combining acoustic and linguistic information have already shown strong performance on the ADReSS and ADReSSo benchmarks: Pan et al.~\cite{pan2025taslp} fused acoustic and language features with cross-attention, and Deng et al.~\cite{deng2022} combined them at the decision level via majority voting, both outperforming single-modality systems. Gu et al.~\cite{gu2026} fused HuBERT speech and RoBERTa transcript via CTC-guided cross-modal alignment and cross-attention, showing that both modalities contribute and that pause and disfluency cues are highly discriminative, and Pu and Zhang~\cite{pu2025pause} showed that encoding pause information into the word embeddings of a language model improves detection. These studies show the benefit of combining complementary information, while the combination strategy remains an important design choice.

However, this diversity creates a problem for conventional ensembling. When classifiers are trained on different representations or with different regularization strengths, their posterior probabilities are not numerically comparable; the same value may correspond to very different confidence levels across models. Simply averaging these probabilities implicitly assumes that the classifier outputs are on comparable probability scales. We refer to this issue as \emph{posterior probability-scale mismatch}.

To address this problem, we propose a rank-based aggregation framework combined with a confidence-gated Random Forest correction stage. The main contributions are summarized as follows:

\begin{itemize}

\item We propose a robust rank aggregation framework that addresses posterior probability-scale mismatch by aggregating normalized prediction ranks instead of posterior probabilities.

\item We introduce a confidence-gated Random Forest that selectively corrects highly confident residual errors using handcrafted linguistic features, providing an interpretable correction stage that identifies which speech dimensions drive the final decision.

\item Experiments on ADReSS2020 and ADReSSo2021 demonstrate that the proposed framework consistently outperforms probability averaging and achieves competitive accuracy on both benchmarks.

\end{itemize}

\begin{figure*}[t]
\centering
\includegraphics[width=0.8\textwidth, page=1]{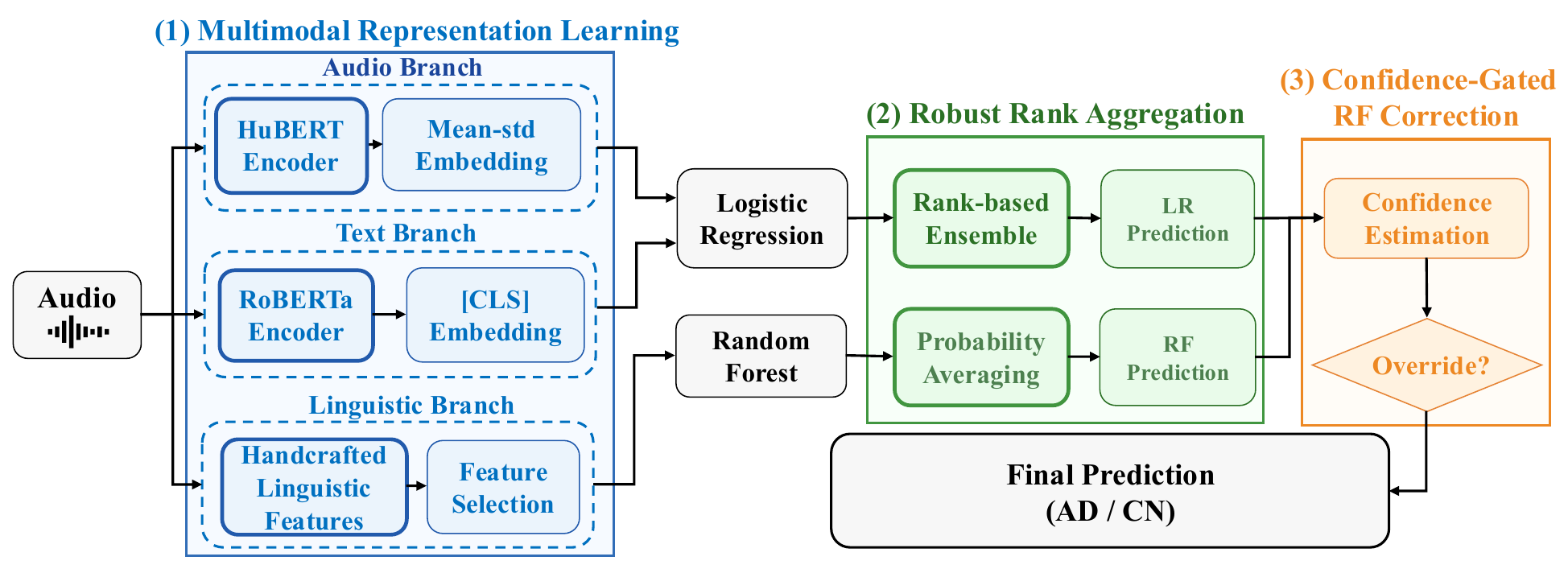}
\caption{Overview of the proposed framework. Three complementary branches extract acoustic, semantic, and handcrafted linguistic representations. Robust rank aggregation mitigates posterior probability-scale mismatch by aggregating normalized prediction ranks, while a confidence-gated Random Forest selectively refines the primary prediction using handcrafted linguistic features.}
\label{fig:framework}
\end{figure*}

\section{Proposed Method}
\label{sec:method}

The proposed framework processes a speech recording and its transcript through three stages: (i) \textbf{representation extraction and classifier outputs}, (ii) \textbf{rank aggregation}, and (iii) \textbf{confidence-gated correction.} Figure~\ref{fig:framework} shows the overall pipeline.

Given a speech recording and its transcript, the framework
extracts acoustic, semantic, and handcrafted linguistic rep-
resentations. These are independently processed by multiple
classifiers whose posterior probabilities may exhibit substan-
tially different numerical scales. Instead of averaging these
probabilities directly, the framework performs robust rank
aggregation to mitigate probability-scale mismatch while pre-
serving classifier confidence ordering. Finally, a confidence-gated RF exploits handcrafted features to selectively correct
highly confident residual errors.
\subsection{Representation and Classifier Outputs}
\label{subsec:representation}

Given an audio recording $A$ and its transcript $X$, we obtain three representations.

\subsubsection*{\textbf{Audio Representation}}

The waveform $A$ is encoded by a frozen HuBERT-large model, a self-supervised speech representation trained without transcripts, which captures non-lexical cues such as prosody, hesitation, and disfluency that complement the text channel. It yields frame-level representations $\{\mathbf{z}_t\}_{t=1}^{T_a}$. The utterance-level audio embedding is
\[
\mathbf{h}_a = [\mathrm{mean}_t(\mathbf{z}_t),\; \mathrm{std}_t(\mathbf{z}_t)].
\]

\subsubsection*{\textbf{Text Representation}}

The transcript $X$ is encoded by a frozen RoBERTa-base model. The sentence-level [CLS] representation is the text embedding $\mathbf{h}_t$.

\subsubsection*{\textbf{Handcrafted Linguistic Features}}

We extract 30 handcrafted acoustic and linguistic features covering pause, timing, content, and lexical information. Specifically, the pool includes
\begin{itemize}
    \item \emph{Pause} features: pause statistics (mean, median, maximum, standard-deviation, and interquartile-range pause duration; pause count, long-pause count, capped pause count, mean capped pause, and pause rate);
    \item \emph{Timing} features: utterance-level timing (number of utterances/turns, total duration, speech rate, mean and standard-deviation utterance duration, words per utterance) and disfluency markers (repetitions and a disfluency-pause interaction);
    \item \emph{Content} features reflecting the semantic richness of the picture description: mentions of people, objects, and actions, together with their ratios (people ratio, objects per utterance, action-to-object ratio) and overall content density;
    \item \emph{Lexical} features quantifying vocabulary and word-level statistics: the type-token ratio and its variants (TTR per utterance, TTR scaled by word count), the number of unique words, and mean and standard-deviation word length.
\end{itemize}

Mutual information (MI) selects the top-$K$ features, yielding the vector $\mathbf{f}$ for the correction stage.

\subsubsection*{Classifier Outputs}

\begin{figure*}[t]
\centering
\begin{minipage}[b]{0.48\textwidth}
\centering
\includegraphics[width=0.84\linewidth]{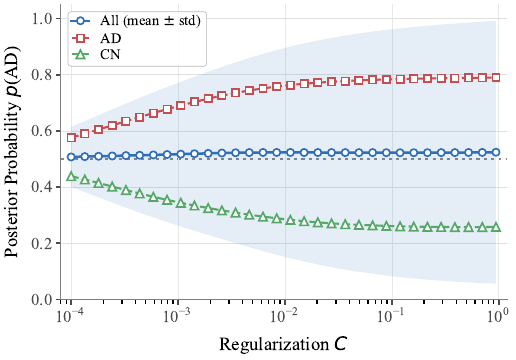}
\caption{Effect of the regularization strength $C$ on posterior probabilities $p(\mathrm{AD})$ for the audio (HuBERT) branch on ADReSS2020.
}
\label{fig:c_prob}
\end{minipage}
\hfill
\begin{minipage}[b]{0.48\textwidth}
\centering
\includegraphics[width=0.94\linewidth, page=2]{figures/pip2_cropped.pdf}
\caption{Comparison of ensemble strategies. Conventional probability averaging directly aggregates posterior probabilities, whereas the proposed method aggregates normalized prediction ranks, making the ensemble robust to posterior probability-scale mismatch.}
\label{fig:rank_ensemble}
\end{minipage}
\end{figure*}

The embeddings $\mathbf{h}_a$ and $\mathbf{h}_t$ are each processed by logistic regression (LR) classifiers trained under different $L_2$ regularization strengths. Let $M$ denote the total number of classifiers, indexed by $j=1,\ldots,M$. For an evaluation sample $i$, classifier $j$ outputs a posterior probability $p_j^{(i)}\in[0,1]$. These probabilities are not directly comparable because the classifiers use different representations and regularization strengths. As shown in Fig.~\ref{fig:c_prob}, changing the regularization strength $C$ alters the spread of the posterior probabilities (the standard deviation across the 48 ADReSS2020 evaluation samples ranges from 0.11 to 0.47) while largely preserving their relative ordering; this motivates aggregating ranks rather than raw probabilities.

\subsection{Rank Aggregation}
\label{subsec:rank_ensemble}

For classifier $j$, we first obtain out-of-fold (OOF) predictions on the training set by leave-one-out cross-validation (LOOCV), denoted $\{q_j^{(n)}\}_{n=1}^{N_{\mathrm{tr}}}$, where $N_{\mathrm{tr}}$ is the number of training subjects. For a test sample $i$, its percentile rank under classifier $j$ is the rank of $p_j^{(i)}$ in the training OOF distribution:

\begin{equation}
\label{eq:rank}
r_j^{(i)}
=
\frac{1}{N_{\mathrm{tr}}}
\sum_{n=1}^{N_{\mathrm{tr}}}
\mathbf1
\!\left(
q_j^{(n)} \le p_j^{(i)}
\right),
\qquad
r_j^{(i)}\in[0,1],
\end{equation}
where $\mathbf1(\cdot)$ is the indicator function. The rank score $r_j^{(i)}$ represents the fraction of training subjects whose OOF score does not exceed $p_j^{(i)}$.

The ensemble rank score is the average over all classifiers:
\[
\bar r^{(i)}
=
\frac{1}{M}
\sum_{j=1}^{M}
r_j^{(i)}.
\]
The rank-based prediction is obtained by thresholding the ensemble rank score at 0.5:
\[
\hat y_{\mathrm{rank}}^{(i)}
=
\mathbf1
\!\left(
\bar r^{(i)} > 0.5
\right).
\]

\subsection{Confidence-Gated RF Correction}
\label{subsec:gated_rf}

A random forest (RF) with $T$ trees is trained on the handcrafted feature vector $\mathbf{f}^{(i)}$. Its probability estimate is
\[
P_{\rm RF}^{(i)}
=
\frac{1}{T}
\sum_{t=1}^{T}
\mathbf1
\!\left(
h_t(\mathbf f^{(i)})=1
\right),
\]
where $h_t(\cdot)$ is the binary prediction of the $t$-th tree. The RF confidence is
\[
c_{\rm RF}^{(i)}
=
2
\left|
P_{\rm RF}^{(i)} - 0.5
\right|.
\]

The RF prediction overrides the rank-based prediction only when they disagree and the RF confidence exceeds a threshold $\tau$. The final prediction is
\[
\hat y^{(i)}
=
\begin{cases}
\hat y_{\rm RF}^{(i)},
&
\hat y_{\rm RF}^{(i)} \neq \hat y_{\rm rank}^{(i)}
\;\;
\text{and}
\;\;
c_{\rm RF}^{(i)} > \tau,\\[1ex]
\hat y_{\rm rank}^{(i)},
&
\text{otherwise},
\end{cases}
\]
where $\hat y_{\rm RF}^{(i)} = \mathbf1(P_{\rm RF}^{(i)} > 0.5)$. This rule limits the influence of the handcrafted branch to high-confidence disagreements, preventing unreliable corrections from degrading the rank ensemble.

The gating threshold $\tau$ is selected on the training side using OOF predictions. We evaluate a range of candidate values for $\tau$ on the training OOF pseudo-ranks and choose the value that maximizes the training OOF accuracy, breaking ties toward the larger $\tau$. This ensures that the RF overrides the rank prediction only for high-confidence disagreements that improve the training-side performance.

\begin{table}[t]
\centering
\caption{ADReSS2020 and ADReSSo2021 test results}
\label{tab:comparison}
\renewcommand{\arraystretch}{0.92}
\footnotesize
\begin{tabular}{lcccc}
\toprule
\textbf{Method} & \textbf{Acc\%} & \textbf{F1\%} & \textbf{Prec\%} & \textbf{Rec\%} \\
\midrule
\multicolumn{5}{l}{\textit{ADReSS2020}} \\
Wang et al.~\cite{wang2022} & 93.75 & 93.90 & 92.00 & \textbf{95.80} \\
Pu and Zhang~\cite{pu2025pause} & 81.20 & 80.90 & 82.60 & 79.20 \\
Gu et al.~\cite{gu2026} & 94.79 & 94.78 & 95.28 & 94.79 \\
Ours & \textbf{95.83} & \textbf{95.65} & \textbf{100.00} & 91.67 \\
\midrule
\multicolumn{5}{l}{\textit{ADReSSo2021}} \\
Deng et al.~\cite{deng2022} & 87.32 & 87.28 & 87.62 & 87.26 \\
Pu and Zhang~\cite{pu2025pause} & 83.10 & 80.60 & \textbf{92.60} & 71.40 \\
Gu et al.~\cite{gu2026} & 88.73 & 88.71 & 88.88 & \textbf{88.69} \\
Ours & \textbf{90.14} & \textbf{89.86} & 91.18 & 88.57 \\
\bottomrule
\end{tabular}

\end{table}

\begin{table}[t]
\centering
\caption{Ablation experiments on ADReSSo2021. $^\dagger$Rank aggregation is replaced by conventional probability averaging.}
\label{tab:ablation}
\renewcommand{\arraystretch}{0.92}
\footnotesize
\begin{tabular}{lcccc}
\toprule
\textbf{Variant} & \textbf{Acc} & \textbf{F1} & \textbf{Prec} & \textbf{Rec} \\
\midrule
Full Pipeline & 90.14 & 89.86 & 91.18 & 88.57 \\
w/o RoBERTa & 83.10 & 81.82 & 87.10 & 77.14 \\
w/o HuBERT & 84.51 & 85.33 & 80.00 & 91.43 \\
w/o Rank Aggregation$^\dagger$ & 85.92 & 85.29 & 87.88 & 82.86 \\
w/o RF Correction & 88.73 & 88.57 & 88.57 & 88.57 \\
\bottomrule
\end{tabular}

\end{table}

\begin{figure*}[t]
\centering
\includegraphics[width=0.9\textwidth]{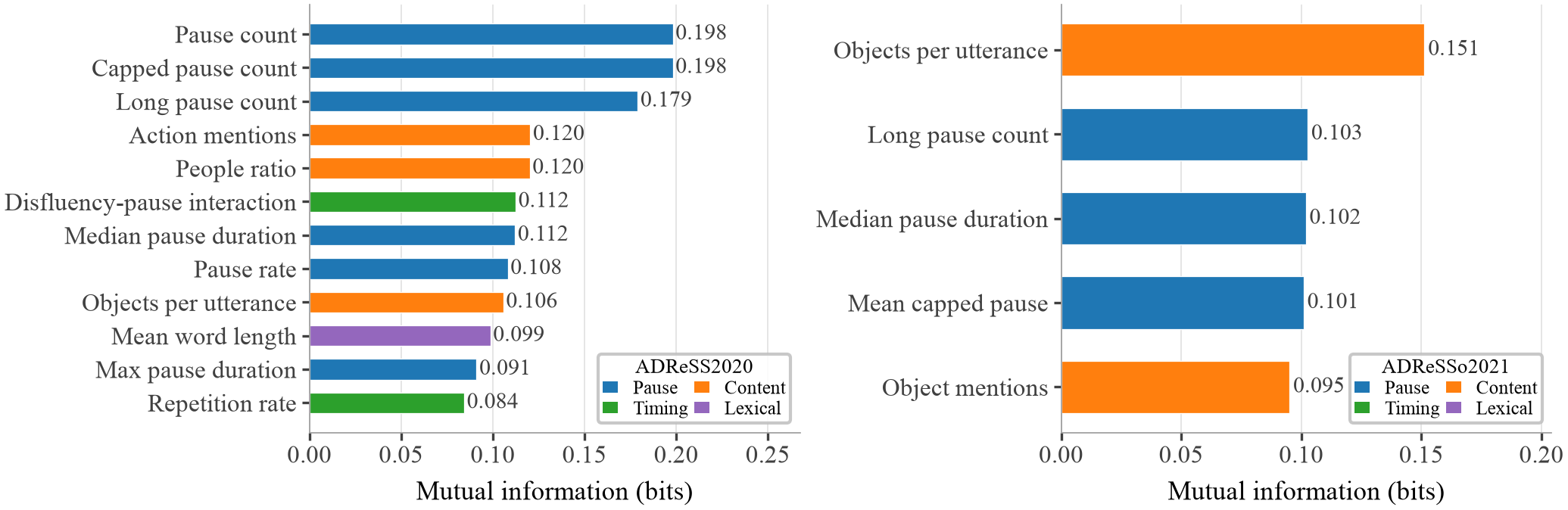}
\caption{MI-selected handcrafted features: top-12 on ADReSS2020 (left) and top-5 on ADReSSo2021 (right). Colors indicate the feature category (Pause/Timing/Content/Lexical).}
\label{fig:featimp}
\end{figure*}

\section{Experiments}
\label{sec:experiments}

\subsection{Experimental Setup}

Experiments are conducted on ADReSS2020 and ADReSSo2021. The audio branch uses frozen HuBERT-large with mean and standard-deviation pooling applied to participant-only speech segments; the text branch uses frozen RoBERTa-base with [CLS] pooling on both datasets. Logistic Regression is used as the base classifier. The ensemble arms are selected from a pool of 66 LR classifiers (33 per modality, trained with different $L_2$ regularization strengths) by forward selection with single- and double-swap refinement on the training rank OOF accuracy, yielding 7 arms for ADReSS2020 and 5 for ADReSSo2021. For the confidence-gated Random Forest, the number of handcrafted features $K$ is chosen from the 30-dimensional pool by maximizing the RF LOOCV OOF accuracy, giving $K{=}12$ for ADReSS2020 and $K{=}5$ for ADReSSo2021, and the gating threshold $\tau$ is selected on the training OOF accuracy as described in Section~\ref{subsec:gated_rf}. All OOF scores are produced by leave-one-out cross-validation (LOOCV).
\subsection{Comparison with Existing Methods}

Table~\ref{tab:comparison} compares the proposed framework with previously reported methods on the ADReSS2020 and ADReSSo2021 benchmarks. The proposed method attains the best accuracy among the compared methods on both datasets. The training-cohort OOF reference achieves 88.57\% F1 on ADReSSo2021, and confidence-gated RF correction improves it to 89.86\% F1. The proposed framework requires no additional deep model training or explicit probability calibration, demonstrating that robust rank aggregation effectively integrates complementary representations.

\subsection{Ablation Study}

Table~\ref{tab:ablation} reports ablations on ADReSSo2021. Removing the RoBERTa text branch causes the largest accuracy drop (7.04), showing that lexical information is indispensable. Removing the HuBERT audio branch causes an accuracy drop of 5.63, confirming the acoustic stream also contributes substantially. Replacing rank aggregation with probability averaging reduces accuracy by 4.22, confirming the benefit of rank-based aggregation. Removing the RF correction costs 1.41, indicating that the gate provides a moderate but consistent improvement. These results show that all components contribute: the text branch is the primary signal, followed by the audio branch, with the gate as a smaller but reliable refinement.

\subsection{Feature Analysis}

Figure~\ref{fig:featimp} shows the top-12 MI-selected handcrafted features for ADReSS2020 and the top-5 for ADReSSo2021. On both datasets, pause and content features occupy the highest ranks, complemented by disfluency and lexical features. These findings are consistent with prior work: encoding pause information into language models improves detection~\cite{pu2025pause}, disfluency and pause features add accuracy over word-only models~\cite{rohanian2021}, linguistic complexity and (dis)fluency features complement pretrained language models~\cite{qiao2021}. Together, the selected features span the pause, timing, content, and lexical categories, supporting the value of the full handcrafted feature set. These selected features also make the correction stage interpretable by identifying which content and pause dimensions drive the RF's decisions. Notably, although the number of selected features differs across the benchmarks, the dominant categories remain the same.

\vspace{-4pt}
\section{Conclusion}
\label{sec:conclusion}

We proposed a rank aggregation framework for multimodal speech-based AD detection. Instead of using classical ensemble methods, like averaging posterior probabilities, we aggregate normalized prediction ranks, making the ensemble invariant to probability-scale mismatch across classifiers trained on different representations. A confidence-gated RF further exploits handcrafted linguistic features for selective error correction, activated only when its prediction is sufficiently confident and disagrees with the primary ensemble.

Ablation studies confirm the contribution of each component: both text and audio modalities are important, and rank aggregation consistently outperforms probability averaging. The outcome of feature selection further shows that the most informative handcrafted features concentrate in the pause and content categories, providing an interpretable correction signal for the final prediction.

Rank aggregation thus offers a principled alternative to probability averaging for classifiers with incomparable output scales. Although demonstrated on Alzheimer's disease detection, the proposed framework is applicable to any multimodal classification problem where heterogeneous classifiers produce incomparable posterior probability scales.

From a practical standpoint, the framework is lightweight, relying on frozen feature extractors and small classifiers, and easy to deploy in clinical screening pipelines. In future work, we plan to extend it to larger clinical cohorts, additional languages, and other heterogeneous classifier families, and to further explore interpretable features selected under the same training-side criterion.

\section{Acknowledgment}
The authors would like to thank Dr. Kai Li for valuable discussions and constructive suggestions on simplifying the proposed method and improving the experimental design.
\bibliographystyle{IEEEbib}
\bibliography{references}

\begin{thebibliography}{10}

\bibitem{who2026}
{World Health Organization},
\newblock ``Dementia,'' \url{https://www.who.int/news-room/fact-sheets/detail/dementia}, 2026,
\newblock Accessed: Jul. 21, 2026.

\bibitem{luz2020}
S.~Luz, F.~Haider, S.~de~la Fuente, D.~Fromm, and B.~MacWhinney,
\newblock ``Alzheimer's dementia recognition through spontaneous speech: The adress challenge,''
\newblock in {\em Proc. INTERSPEECH}, 2020, pp. 2172--2176.

\bibitem{luz2021}
S.~Luz, F.~Haider, S.~de~la Fuente, D.~Fromm, and B.~MacWhinney,
\newblock ``Detecting cognitive decline using speech only: The adresso challenge,''
\newblock in {\em Proc. INTERSPEECH}, 2021, pp. 3780--3784.

\bibitem{yuan2020}
J.~Yuan, Y.~Bian, X.~Cai, J.~Huang, Z.~Ye, and K.~Church,
\newblock ``Disfluencies and fine-tuning pre-trained language models for detection of alzheimer's disease,''
\newblock in {\em Proc. INTERSPEECH}, 2020, pp. 2162--2166.

\bibitem{qiao2021}
Y.~Qiao, X.~Yin, D.~Wiechmann, and E.~Kerz,
\newblock ``Alzheimer's disease detection from spontaneous speech through combining linguistic complexity and (dis)fluency features with pretrained language models,''
\newblock in {\em Proc. INTERSPEECH}, 2021, pp. 3805--3809.

\bibitem{krogh1994}
A.~Krogh and J.~Vedelsby,
\newblock ``Neural network ensembles, cross validation, and active learning,''
\newblock in {\em Advances in Neural Information Processing Systems (NeurIPS)}, 1994, vol.~7.

\bibitem{hsu2021}
W.-N. Hsu, B.~Bolte, Y.-H.~H. Tsai, K.~Lakhotia, R.~Salakhutdinov, and A.~Mohamed,
\newblock ``{HuBERT}: Self-supervised speech representation learning by masked prediction of hidden units,''
\newblock {\em IEEE/ACM Trans. Audio, Speech, Language Process.}, vol. 29, pp. 3451--3460, 2021.

\bibitem{liu2019}
Y.~Liu, M.~Ott, N.~Goyal, J.~Du, M.~Joshi, D.~Chen, O.~Levy, M.~Lewis, L.~Zettlemoyer, and V.~Stoyanov,
\newblock ``{RoBERTa}: A robustly optimized {BERT} pretraining approach,''
\newblock {\em arXiv:1907.11692}, 2019.

\bibitem{pan2025taslp}
Y.~Pan, B.~Mirheidari, D.~Blackburn, and H.~Christensen,
\newblock ``A two-step attention-based feature combination cross-attention system for speech-based dementia detection,''
\newblock in {\em IEEE/ACM Trans. Audio, Speech, Lang. Process.}, 2025, vol.~33, pp. 896--907.

\bibitem{deng2022}
H.~Deng, H.~Liu, Y.~Zhou, and G.~Lu,
\newblock ``Alzheimer's disease detection using acoustic and linguistic features,''
\newblock in {\em Proc. IEEE HPCC/DSS/SmartCity/DependSys}, 2022, pp. 2280--2284.

\bibitem{gu2026}
M.~Gu, Z.~Tan, K.~Li, X.~Wang, B.~Wen, T.~Wang, G.~Zhang, and J.~Dang,
\newblock ``Interval-aware retrieval framework for speech-based automatic alzheimer's detection,''
\newblock in {\em Proc. ICASSP}, 2026, pp. 8102--8106.

\bibitem{pu2025pause}
Y.~Pu and W.-Q. Zhang,
\newblock ``Integrating pause information with word embeddings in language models for alzheimer's disease detection from spontaneous speech,''
\newblock in {\em Proc. ICASSP}, 2025, pp. 1--5.

\bibitem{wang2022}
Y.~Wang, T.~Wang, Z.~Ye, L.~Meng, S.~Hu, X.~Wu, X.~Liu, and H.~M. Meng,
\newblock ``Exploring linguistic feature and model combination for speech recognition based automatic ad detection,''
\newblock in {\em Proc. INTERSPEECH}, 2022, pp. 3328--3332.

\bibitem{rohanian2021}
M.~Rohanian, J.~Hough, and M.~Purver,
\newblock ``Alzheimer's dementia recognition using acoustic, lexical, disfluency and speech pause features robust to noisy inputs,''
\newblock in {\em Proc. INTERSPEECH}, 2021, pp. 3820--3824.

\end{thebibliography}

\end{document}